\documentclass[aps,pra,superscriptaddress,showpacs,showkeys,preprint]{revtex4}
\usepackage{epsfig}
\usepackage{graphicx}
\usepackage{bm}
\usepackage{amssymb}
\newcommand{\zh}{\bm}
\newcommand{\real}{\mathop{\rm Re}\nolimits}

\newcommand{\Ordnung}{\mathop{\rm O}\nolimits}

\newcommand{\oee}{{\epsilon}}
\newcommand{\mee}{{E}}

\newcommand{\dee}{{\varepsilon}}

\newcommand{\toee}{{\varepsilon}}
\newcommand{\jj}{{\mbox{$j$--$j$} }}
\newcommand{\eg}{{\it e.g.}}
\newcommand{\ie}{{\it i.e.}}

\newcommand{\ini}{{i}}
\newcommand{\fin}{{f}}

\newcommand{\Ini}{{I}}
\newcommand{\Fin}{{F}}

\newcommand{\zhr}{{\zh r}}

\newcommand{\zhp}{{\zh p}}
\newcommand{\zhk}{{\zh k}}

\newcommand{\zhnu}{{\zh\nu}}
\newcommand{\zhA}{{\zh A}}
\newcommand{\hzhk}{{\zhnu_k}}
\newcommand{\hzhp}{{\zhnu_p}}

\newcommand{\setg}{{g}}
\newcommand{\bpsi}{{\bar{\psi}}}

\newcommand{\Br}[1]{(\ref{#1})}
\newcommand{\Eq}[1]{Eq.\ (\ref{#1})}

\newcommand{\Fig}[1]{Fig.\ \ref{#1}}
\newcommand{\txt}[1]{{\rm #1}}

\newcommand{\um}[1]{\overline{\vphantom{I}#1}}
\newcommand{\xmatrix}[4]{
        \left[
        \begin{array}{cc}
        #1  & #2  \\
        #3  & #4  \\
        \end{array}
        \right]
        }
\newcommand{\threej}[6]{
        \left(
        \begin{array}{ccc}
        #1  & #2  & #3 \\
        #4  & #5  & #6 \\
        \end{array}
        \right)
        }
\newcommand{\sixj}[6]{
        \left\{
        \begin{array}{ccc}
        #1  & #2  & #3 \\
        #4  & #5  & #6 \\
        \end{array}
        \right\}
        }

\begin{document}

\title{Radiative double electron capture by bare nucleus with emission of one photon}

\author{E.\ A.\ Chernovskaya}
\email{evgenia.chernovskaya@yandex.ru}
\affiliation{Faculty of Physics,
             St.\ Petersburg State University, Ulyanovskaya 1,
             198504,
             Petergof, St.\ Petersburg, Russia}
\author{O.\ Yu.\ Andreev}
\email{olyuan@gmail.com}
\affiliation{Faculty of Physics,
             St.\ Petersburg State University, Ulyanovskaya 1,
             198504,
             Petergof, St.\ Petersburg, Russia}
\author{L.\ N.\ Labzowsky}
\affiliation{Faculty of Physics,
             St.\ Petersburg State University, Ulyanovskaya 1,
             198504,
             Petergof, St.\ Petersburg, Russia}
\affiliation{Petersburg Nuclear Physics Institute,
             188300,
             Gatchina, St.\ Petersburg, Russia}

\date{\today}

\begin{abstract}
Calculation of the cross-section for the process of double electron capture by bare nucleus with emission of a single photon is presented. The double electron capture is evaluated within the framework of Quantum Electrodynamics (QED).
Line-Profile Approach (LPA) is employed.
Since the radiative double electron capture is governed by the electron correlation, corrections to the interelectron interaction were calculated with high accuracy, partly to all orders of the perturbation theory.
\end{abstract}


\maketitle

%
\section{Introduction}
\label{introduction}
The processes of electron capture are under intensive investigation of the experimentalists and theoreticians during the last decades.
Still there are some disagreements between the experimental and theoretical results.

\par
One of the observed processes is the process of Radiative Electron Capture (REC) accompanied by the emission of one photon.
There are many experimental data available
\cite{stohlker95p2098,eichler07}.
REC is the dominant electron-capture channel in fast collisions of heavy ions with light target atoms.
This process does not depend strongly on the interelectron interaction.

\par
Interelectron interaction becomes important in the process of the capture of two electrons.
There are two different types of the processes with the capture of two electrons by the Highly Charged Ions(HCI): Double Radiative Electron Capture (DREC) and Radiative Double Electron Capture (RDEC).
DREC is a two-step process in which two uncorrelated electrons are captured in one collision and two photons are emitted.
RDEC is a one-step process where the momentum and the energy of two correlated captured electrons are converted into energy and momentum of one emitted photon.
The processes of double electron capture were investigated experimentally in
\cite{warczak95,bednarz01,bednarz03,simon10}
and theoretically in
\cite{nefiodov05,yakhontov97,mikhailov04p350,voitkiv06}.

\par
In experiments
\cite{warczak95,bednarz01,bednarz03,simon10}
RDEC was organized as a process where two free (or quasifree) target electrons are captured into a bound state of the projectile, e.g., into empty $K$-shell of an ion, and the energy emerges with only one photon.
This process can be treated as the inverse process to the double photoionization.
But in RDEC unlike the photoionization the bare nuclei should be used.
The RDEC is a convenient tool to investigate the electron-electron interaction in the processes of the ion-atom collisions.

\par
The first RDEC experiment
\cite{warczak95}
was performed with $11.4$~MeV/u Ar${}^{18+}$ ions hitting a carbon foil at the UNILAC in GSI in 1994.
To obtain as much as possible high rate of double capture in one collision, solid carbon target was chosen.
In this experiment the probability of the RDEC process is very small.
Experimental data give only the upper limit of the value of the cross-section of the process: about $5.2$~mb.

\par
Another RDEC experiment
\cite{bednarz03}
was performed at the heavy ion storage ring (ESR) in GSI.
Bare U${}^{92+}$ ions with the energy of $297$~MeV/u have been used in collisions with gas of Ar atoms.
From the data obtained in this experiment we can only conclude that the value of the cross-section is also very small: less than $10$~mb.
The upper limit of RDEC process was estimated to be significantly lower than measured previously
\cite{bednarz01}.

\par
Recent experiment
\cite{simon10}
with $38$~MeV O${}^{8+}$ ions shows that there is some discrepancy between the experimental data and theoretical prediction.
In this experiment ions of the oxygen O${}^{8+}$ move trough the thin carbon foil.

\par
The RDEC process was investigated theoretically in
\cite{yakhontov97,voitkiv06,mikhailov04p350,nefiodov05}.
In these works the calculations were performed within the nonrelativistic theory.

\par
The process of double electron capture is governed by the interelectron interaction.
We investigate radiative double electron capture (RDEC) by a bare nucleus followed by emission of a single photon
\begin{eqnarray}
2e^{-}(\epsilon) + X^{(Z)+}
&\to&
X^{(Z-2)+}(1s1s) + \gamma(\omega)
\,.
\end{eqnarray}
The initial state is presented by two incident electrons $2e^{-}$ with the same energies $\epsilon$ and a bare atomic nucleus $X^{(Z)+}$ with the charge $Z$.

\par
The final state is given by two-electron ion in the ground state $X^{(Z-2)+}(1s1s)$ and a single photon $\gamma(\omega)$.
We concern only radiation of one photon, first, because in the experiment
\cite{simon10}
one photon is registered, and second, since the radiation of one photon gives the major contribution to the value of cross-section of the process.

\par
We consider a scenario where the momenta of the incident electrons are the same for the both electrons.
This scenario corresponds to the experimental situation
\cite{warczak95,simon10,bednarz03}.
The results of our calculations are compared with available experimental and theoretical data.

%
\section{Application of the Line-Profile Approach}
\label{formalism}
The double electron capture is a nonresonant process.
However for its description we will apply the LPA first developed for the resonant processes.
The LPA appeared to be a convenient tool for the description of the nonresonant processes as well.

\par
Within the framework of quantum electrodynamics (QED) atomic electrons are interacting with the quantized electromagnetic field and with the quantum vacuum
\cite{akhiezer65b}.
Accordingly, the set of electrons (together with the atomic nucleus) is not a conservative system, and the concept of the energy for this system needs to be carefully treated.
Within the LPA
\cite{andreev08pr}
the energy levels are associated with resonances in the natural line-profile for the process of resonant photon scattering.
In order to keep the characteristics of the energy levels independent on the particular features of the process of scattering, the resonance approximation is employed.
The resonance approximation consists of description of resonance area of the natural line-profile by the Lorentz contour which is characterized by two parameters: position of the resonance and its width.
The energy levels are connected with the corresponding resonances.
The energy and width of an energy level are determined by position of the resonance and its width within the resonance approximation.

\par
The amplitude of the process of photon scattering (the initial and final states are assumed to be the ground state) can be presented as a matrix element of a special operator
\cite{andreev01,andreev08pr}.
This operator can be constructed with employment of the QED perturbation theory.
In general, this operator depends on the photon frequency ($\omega$) and can be considered as a complex symmetric matrix (in some basis set) or as a quadratic form reducible to a diagonal form.
Within the resonance approximation the eigenvalues of this matrix determine the positions of resonances and their widths.
The eigenvectors of this matrix are used for calculation of the transition probabilities.
If we consider the probability of a particular transition between two energy levels, we need to calculate the amplitude of this process.
The amplitude is derived as a matrix element of the photon emission (absorption) operator (also constructed with employment of the QED perturbation theory) calculated on the eigenvectors corresponding to the initial and final states.
Application of the LPA to the evaluation of energies and transition probabilities is presented in detail in
\cite{andreev08pr,andreev01,andreev09p032515}.
We note that the technique developed in
\cite{andreev08pr,andreev01,andreev09p032515}
can be used only for the bound electrons.

\par
The aim of the present work is the evaluation of the cross-section for electron capture.
This process can be considered as a transition
\begin{eqnarray}
\Ini
&\to&\label{eq080823n01}
\Fin
\,,
\end{eqnarray}
where the initial state ($\Ini$) in case of REC process corresponds to the two electrons: a bound $1s$-electron and an incident electron, \ie, continuum electron.
The final state ($\Fin$) is represented by two bound electrons in the \jj coupling scheme configuration.
In case of RDEC the initial state corresponds to the two continuum electrons and the final state is the same.
Since the initial state contains continuum electrons, the LPA can not be applied directly to these processes.
However, we can introduce an auxiliary bound electron system which properties are explicitly related to the properties of original system.

\par
We can consider the highly charged ion being confined within a sphere of a large radius $R$.
Then, all the energy spectrum becomes discrete.
If the radius of the sphere is large, the wave function of election confined within the sphere (spherical box) and the wave function of electron (with the same energy) not confined within the sphere almost coincide.
Let the electrons have the energy $\epsilon > m_e c^2$.
Eigenvectors of the Dirac equation for the point nucleus are well known
\cite{akhiezer65b}.
The asymptotics ($r\to\infty$) of the Dirac wave function for the electron in continuum reads
\begin{eqnarray}
\psi_{\epsilon jlm}(\zhr)
&=&\label{eq080824n01}
\frac{1}{r}
\left({{g_{\epsilon}(r)\,\Omega_{jlm}(\zhnu_r)}\atop{i f_{\epsilon}(r)}\,\Omega_{j,2j-l,m}(\zhnu_r)}\right)
\\
g_{\epsilon}(r)
&=&\label{eq080823n05}
C_g
\sqrt{\frac{\oee+m}{\pi p}}
\cos(pr+\phi_g(r))
\\
f_{\epsilon}(r)
&=&\label{eq080823n06}
C_f
\sqrt{\frac{\oee-m}{\pi p}}
\sin(pr+\phi_f(r))
\,,
\end{eqnarray}
where $|C_g|=|C_f|=1$ and $\phi_g(r)$, $\phi_f(r)$ are the functions smoothly depending on $r=|\zhr|$, $\zhnu_r=\zhr/r$.
The energy ($\epsilon$) and momentum ($p$) of the electron are connected as $\epsilon^2=m_e^2c^4+p^2$, where $m_e$ is the electron mass, $c$ is the speed of light.
The continuum electron function is normalized to the energy delta function.
For the large radius $R$ and coordinate $r$ the electron wave function for the ion enclosed within a box is given by its asymptotic:
Eqs.\ \Br{eq080823n05}, \Br{eq080823n06}.
Accordingly, the difference between the nearest (discrete) values of the momentum ($\Delta p$) is defined by one half of the oscillation period of functions in
Eqs.\ \Br{eq080823n05}, \Br{eq080823n06} at the border ($r=R$):
$\Delta p R=\pi$.
Then, the difference between the nearest values of the energy ($\oee$) is
\begin{eqnarray}
\Delta \oee
&=&\label{eq080823n07}
\frac{p}{\oee}\Delta p
=
\frac{p\pi}{\oee R}
\,.
\end{eqnarray}
The equations in this Section should be understood in the asymptotic sense, \ie, the equations are correct up to the terms disappearing when $R\to\infty$.

\par
The wave function of electron confined within the sphere of radius $R$ can be written as
\cite{akhiezer65b}
\begin{eqnarray}
\psi_{e_R}(\zhr)
&=&\label{eq110403n04}
\frac{1}{N_{e_R}}\psi_{e}(\zhr)\,\theta(R-|\zhr|)
\,,
\end{eqnarray}
where $\psi_{e}(\zhr)$ is given by
\Eq{eq080824n01},
\begin{eqnarray}
(N_{e_R})^2
&=&\label{eq110403n02}
\frac{\epsilon R}{p\pi}
=
\frac{1}{\Delta\epsilon}
\,
\end{eqnarray}
is a normalization factor (($\psi_{e_R}(\zhr)$ is normalized to unity), $\theta(R-|\zhr|)$ is the Heaviside step function.
The normalization factor $N_{e_R}$ is considered in detail in
Appendix~\ref{appendixb}
(see
\Eq{eqn111016n20}).
Note, that the function $\psi_{e_R}(\zhr)$ goes to zero at any $\zhr$ when $R\to\infty$ as
\begin{eqnarray}
\psi_{e_R}(\zhr)
&\sim&\label{eq080825n03}
\frac{1}{\sqrt{R}}
\,.
\end{eqnarray}

\par
At a fixed radius $R$ the function $\psi_{e_R}(\zhr)$ describes a discrete energy level.
Technically, we can consider a resonance corresponding to this level.
So, the LPA can be applied to any energy level of the HCI confined within a sphere of a finite radius.

\par
Reasoning by analogy with the system not confined within a sphere, \ie, with a system which has a continuous part of energy spectrum
\cite{akhiezer65b},
instead of consideration of a single energy level ($e_R$), we have to consider all the energy levels within some interval $\delta\epsilon=[\oee_1,\oee_2]$.
The number of levels within this interval is proportional to $1/R$ (see
\Eq{eq080823n07}).
The integration over an interval $[\oee_1,\oee_2]$ in the continuous spectrum is equivalent to the summation over all the states ($n$) with the energy ($\oee_n$) from the interval $[\oee_1,\oee_2]$ in the discrete spectrum (if the ion is enclosed within a sphere of radius $R$):
\begin{eqnarray}
\int\nolimits^{\oee_2}_{\oee_1}d\oee'\,
F(\oee')
&=&
\sum\limits_{\oee_n\in[\oee_1,\oee_2]}
F(n)
\,,
\end{eqnarray}
where function $F$ represents some physical property (\eg, cross-section).
If the radius $R$ goes to infinity, the number of discrete states in the energy interval $[\oee_1,\oee_2]$ goes to infinity and the width of the energy interval containing only one state goes to zero.
Accordingly, if $\delta\oee\to\Delta\oee$, we can write
\begin{eqnarray}
F(\oee)
&=&
\frac{1}{\Delta \oee}
\int\limits_{\Delta \oee}d\oee'\,
F(\oee')
=
\frac{1}{\Delta \oee}
F(n)
=
(N_{e_R})^2
F(n)
\,,
\end{eqnarray}
where $\oee_n=\oee$ is the only discrete state inside the energy interval $\Delta\oee$.
Thus, the transformation from the continuous to discrete spectrum results in
the substitution of the continuous spectrum wave function $\psi_e$ by the function $\psi_{e_R}$ and in an additional factor $1/\Delta\oee=(N_{e_R})^2$ to the function $F$ (cross-section), where $\Delta\oee$ is the distance between the nearest energy levels.

\par
We conclude that the LPA can be generalized to the case of continuum electrons in the initial or final states.
We can introduce an artificial bound electron state $e_R$ described by the wave function $\psi_{e_R}$.
The energies and the angular quantum numbers of the continuum electron state $\psi_{e}$ and the bound electron state $\psi_{e_R}$ are equal.
If there is one continuum electron in the initial or final states, the amplitude calculated with functions $\psi_{e}$ and the amplitude calculated with functions $\psi_{e_R}$ are related like
\cite{andreev09p042514}
\begin{eqnarray}
U_{e}
&=&\label{eq080823n02}
\lim\limits_{R\to\infty}
N_{e_R}
U_{e_R}
\,.
\end{eqnarray}
If there are two continuum electron in the initial or final states, the amplitudes are related like
\begin{eqnarray}
U_{e_1 e_2}
&=&\label{eq080823n02x2}
\lim\limits_{R\to\infty}
N_{{e_1}_R}N_{{e_2}_R}
U_{{e_1}_R {e_2}_R}
\,,
\end{eqnarray}
where $N_{{e_i}_R}$ is the normalization constant for the corresponding electron given by
\Eq{eq110403n04}.
In this paper we will consider electrons with equal energies, accordingly, we can set $N_{{e}_R}\equiv N_{{e_1}_R}=N_{{e_2}_R}$.
We note, that the limit $R\to\infty$ is equivalent to limit $N_{\oee_R}\to\infty$.

\par
So, we can generalize the LPA for calculation of the amplitude of the process of the electron capture.
We employ the artificial bound electron states $e_R$ defined by
\Eq{eq110403n04}
and apply the LPA for calculation of the transition amplitude ($U_R$), \ie, for the system where the continuum electrons are substituted by the bound electrons $e_R$.
The amplitude of the electron capture is given by
\Eq{eq080823n02}
(if there is one continuum electron)
or by
\Eq{eq080823n02x2}
(if there are two continuum electrons).
The limit $R\to\infty$ can be evaluated numerically.

\section{Two-electron wave functions}
\label{dcontinuum}
\par
The incident electron can be characterized by momentum ($\zhp$) and polarization or spin projection ($\mu$), and described by wave function $\psi_{\zhp\mu}(\zhr)$.
The energy ($\oee$), momentum and electron mass ($m_e$) are connected as $\oee=\sqrt{p^2 + m_e^2}$, where $p=|\zhp|$.
It is also convenient to introduce the electron wave vector $\hzhp=\zhp/|\zhp|$.
The wave function of incident electron is normalized like
\begin{eqnarray}
\int d\zhr\,\psi^{+}_{\zhp'\mu'}(\zhr)\psi_{\zhp\mu}(\zhr)
&=&\label{eqn090514n02}
(2\pi)^3 \delta(\zhp'-\zhp)\delta_{\mu'\mu}
\\
&=&
\frac{(2\pi)^3}{p\oee}
\delta(\oee'-\oee)
\delta(\cos{\theta'}-\cos{\theta})\delta(\phi'-\phi)\delta_{\mu'\mu}
\,,
\end{eqnarray}
where the set $(p,\theta,\phi)$ represents the vector $\zhp$ in spherical coordinates.
This normalization corresponds to one particle per unit volume.

\par
The wave function of the incident electron ($\psi_{\zhp\mu}$) can be expanded in the complete set of wave functions ($\psi_{\toee jl m}$) with the certain energy ($\toee$), total angular momentum ($j$), parity ($l$) and projection of the total angular momentum ($m$)
\cite{akhiezer65b}
\begin{eqnarray}
\psi_{\zhp\mu}(\zhr)
&=&\label{eqn090514n01}
\int d\toee\,
\sum_{jl m}
a_{\zhp\mu,\toee jlm}
\psi_{\toee jl m}(\zhr)
\,.
\end{eqnarray}
Introducing the scalar product
\begin{eqnarray}
\alpha_{\hzhp\mu,jlm}
&=&\label{eqn110919n6}
(\Omega^{+}_{jlm}(\hzhp)v^{\mu}(\hzhp))
\,,
\end{eqnarray}
where $\Omega_{jlm}(\hzhp)$ is the spherical spinor
\cite{varshalovich}
and $v^{\mu}(\hzhp)$ is the unit spinor function, the scalar products $a_{\zhp\mu,\oee' jlm}$ can be presented in the form
\begin{eqnarray}
a_{\zhp\mu,\toee jlm}
&=&\label{eqn110919n1}
\frac{(2\pi)^{3/2}}{\sqrt{p\oee}}
\delta(\toee-\oee)
e^{i\varphi_{\dee jl}}
\alpha_{\hzhp\mu,jlm}
\,.
\end{eqnarray}
The phase $\varphi_{\dee jl}$ is determined by the field of the nucleus
\cite{akhiezer65b}.
The functions $\psi_{\toee jlm}(\zhr)$ are normalized like
\begin{eqnarray}
\int d\zhr\, \psi^{+}_{\toee'j'l'm'}(\zhr) \psi_{\toee jlm}(\zhr)
&=&\label{eqn110622}
\delta(\toee'-\toee)
\delta_{j'j}\delta_{l'l}\delta_{m'm}
\,.
\end{eqnarray}

\par
The wave function describing two incident electrons with the certain momenta and polarizations can be written as
\begin{eqnarray}
\Psi_{\zhp_1\mu_1,\zhp_2\mu_2}(\zhr_1,\zhr_2)
&=&
\frac{1}{\sqrt{2}}
\det\{ \psi_{\zhp_1\mu_1}(\zhr_1),\psi_{\zhp_2\mu_1}(\zhr_2)\}
\,.
\end{eqnarray}

\par
We suppose that the initial state of the system is given by two incident electrons with the equal momenta ($\zhp$) and the opposite polarizations ($\mu_1=-\mu_2$).
Accordingly, the wave function of the initial state is
\begin{eqnarray}
\Psi^\txt{ini}(\zhr_1,\zhr_2)
&=&\label{eqn110919n4}
\frac{1}{\sqrt{2}}
\det\{ \psi_{\zhp \mu_1=1/2}(\zhr_1),\psi_{\zhp \mu_2=-1/2}(\zhr_2)\}
\,.
\end{eqnarray}

%
\section{cross-section}
\label{electron}
The amplitude of the process of electron capture ($U_{\ini\fin}$) is defined via S-matrix
\cite{akhiezer65b}
\begin{eqnarray}
S_{\ini\fin}
&=&
(-2\pi i)\delta(\mee_\fin-\mee_\ini)U_{\ini\fin}
\,,
\end{eqnarray}
where $\mee_\ini$, $\mee_\fin$ are the energies of the initial and final states of the system. 
Then, the transition probability is given by
\cite{bogoliubov80b}
\begin{eqnarray}
dw_{\ini\fin}
&=&\label{eqn090507n01}
2\pi 
\frac{1}{V^2}
|U_{\ini\fin}|^2
\delta(\mee_\fin-\mee_\ini)
\frac{d\zhk}{(2\pi)^3}
\,,
\end{eqnarray}
where $\mee_\ini$, $\mee_\fin$ are the initial and final energies of the whole system.
The factor $1/V^2$ corresponds to the densities of the incident electrons, $V$ is the reaction volume.
The wave functions of the incident electrons are normalized by the condition: one particle per unit volume.
The factor $d\zhk/(2\pi)^3$ gives the number of photon states with certain polarization and momentum within an interval $d\zhk$ per unit volume: $d\zhk/(n^\txt{ph}(2\pi)^3)$, $n^\txt{ph}$ is the photon density.
The emitted photon is described by momentum ($\zhk$), frequency ($\omega=|\zhk|$) and polarization ($\lambda$).
Normalization of the photon wave function ($A=(A^{0},\zhA)$), corresponding to one particle per unit volume, is
\begin{eqnarray}
\int d\zhr A^{(\zhk,\lambda)+}(\zhr)A^{(\zhk',\lambda')}(\zhr)
&=&\label{photon-wavefunction-norm}
(2\pi)^{3}
\frac{4\pi}{2\omega}
\delta(\zhk-\zhk')
\delta_{\lambda,\lambda'}
\,.
\end{eqnarray}
Accordingly, the photon density ($n^\txt{ph}$) is set equal to unity.

\par
Cross-section is connected with the transition probability
\Br{eqn090507n01}
as
\cite{akhiezer65b}
\begin{eqnarray}
d\sigma_{\ini\fin}
&=&\label{eqn111016n21}
\frac{dw_{\ini\fin}}{j}
\,,
\end{eqnarray}
where $j$ is the current of the incident electrons.
This current is defined as $j=n^\txt{e}v$, where $n^\txt{e}=1/V$ and $v=p/\oee$ are the density and velocity of the incident electrons, respectively, in the rest system of the nucleus.

\par
\begin{figure}
\centerline{\mbox{\epsfysize=100pt \epsffile{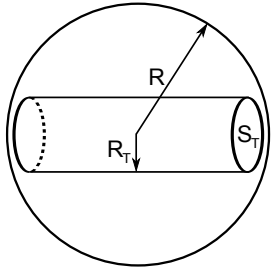}}}
\caption{
In the experiments the bare nucleus is moving through fixed target atoms.
The present calculation is performed in the rest frame of the bare nucleus.
The system is enclose into sphere of the radius $R$.
The bare nucleus is fixed in the center of the sphere.
The cylinder presents the reaction volume for the process of radiative double electron capture.
The area of the cross-section of the cylinder ($S_\txt{T}=\pi R_\txt{T}^2$) is given by the radius of the target atom ($R_\txt{T}$).
The volume of the cylinder is $V_\txt{T}=2RS_\txt{T}$.
The reaction volume for one electron is $V=V_\txt{T}/Z_\txt{T}$, where $Z_\txt{T}$ is the number of electron in the target atom.
}
\label{fig-cylinder}
\end{figure}

In the experiments
\cite{warczak95,simon10,bednarz01,bednarz03}
the RDEC is considered as a process where a bare nucleus goes through target atoms and captures two electrons with emission of one photon.
In our theoretical model this process is considered in the rest frame of the bare nucleus.
Accordingly, the incident electrons are located in the target atom.
The reaction volume for one incident electron is
\begin{eqnarray}
V
&=&
\frac{V_\txt{T}}{Z_\txt{T}}
\,,
\end{eqnarray}
where $V_\txt{T}$ is the reaction volume the target atom, $Z_\txt{T}$ is the number of electrons in the target atom.
We introduce the reaction volume $V$; within this volume the incident electron interacts with the nucleus.
If the system is enclosed into sphere of a large radius $R$, then the reaction volume for the target atom (see
Fig. \ref{fig-cylinder})
is represented by cylinder which cross-section area is $S_\txt{T}=\pi R_\txt{T}^2$ ($R_\txt{T}$ is the target atom radius) and the length is equal to $2R$: $V_\txt{T}=2RS_\txt{T}$.
The reaction volume for one incident electron is
\begin{eqnarray}
V
&=&
2RS
\,,
\end{eqnarray}
where $S$ is the area of the cross-section of the reaction volume for one incident elctron: $S={S_\txt{T}}/{Z_\txt{T}}$.
The volume $V$ can be expressed via the normalization constant $N_{e_R}$ (see
\Eq{eq110403n02})
\begin{eqnarray}
V
&=&\label{eqn110919n2}
\frac{2\pi p N_{e_R}^2  S}{\oee}
\,.
\end{eqnarray}
Here, we took into account that the incident electrons have the same energy, and therefore their normalization constants ($N_{e_R}$) are equal.

\par
In this work we calculate the total cross-section of the electron recombination, what means the integration over the directions of the emitted photon ($\hzhk$) and summation over the photon polarization ($\lambda$).
Then, we suppose that the incident electrons have the same momentum ($\hzhp$), hence, we can also average over the electron momentum direction ($\hzhp$).

\par
It is convenient to make a decomposition
\Eq{eqn090514n01}
of the continuum electron wave function with certain momentum ($\zhp$) and polarization ($\mu$) over the electron wave functions with certain energy ($\toee$), total angular momentum ($j$), parity ($l$) and projection of the total angular momentum ($m$).
The initial state (two incident electrons) is decomposed over the two-electron functions in the \jj coupling scheme.

\par
Accordingly, the cross-section can be written down as
\cite{andreev09p042514}
\begin{eqnarray}
\sigma_{\ini\fin}
&=&\label{eqn090514n13}
\lim\limits_{N_{\oee_R}\to\infty}
\frac{\omega^2}{(2\pi)^2}
\left[
\frac{\oee}{p}
\frac{1}{4\pi}
\,
N_{\oee_R}^4
\,
\frac{\oee}{2\pi p N_{\oee_R}^2 S}
\right]
\int d\hzhk
\, d\hzhp
\left|U_{i,\zhk\lambda s}\right|^2
\end{eqnarray}
where the photon frequency ($\omega$) is defined by the energy conservation law.
The factor $\oee/p$ in the square brackets comes from the current of the incident electrons.
The factor $1/4\pi$ represents the average over the direction of the momentum of the incident electrons ($\hzhp$); we suppose that the momenta of the incident electrons are equal.
The factor $N_{\oee_R}^4$, according to
\Eq{eq080823n02x2},
shows that in the amplitude the one-electron wave functions ($\psi_{\oee_R jlm}$) are normalized to unity (see
\Eq{eq110403n04}).
The last factor in the square brackets is the contribution of the volume given by
\Eq{eqn110919n2}.
The first subindex ($i$) of the amplitude represents the initial state
\Eq{eqn110919n4}.
The other sub-indices represent the final state:
the sub-indices $\zhk \lambda$ describe the emitted photon,
the subindex $s=(J_sM_sn_{s_1}j_{s_1}l_{s_1}n_{s_2}j_{s_2}l_{s_2})$ corresponds to the two-electron configuration in the \jj coupling scheme
\begin{eqnarray}
\Psi_{JMn_{1}j_{1}l_{1}n_{2}j_{2}l_{2}}(\zhr_1,\zhr_2)
&=&\nonumber
N
\sum\limits_{m_1m_2}
C^{j_{1}j_{2}}_{JM}(m_1,m_2)
\\
&&\label{eqn090514n10}
\times
(\psi_{n_{1}j_{1}l_{1}m_1}(\zhr_1)\psi_{n_{2}j_{2}l_{2}m_2}(\zhr_2)-
\psi_{n_{2}j_{2}l_{2}m_2}(\zhr_1)\psi_{n_{1}j_{1}l_{1}m_1}(\zhr_2))
\,,
\end{eqnarray}
where $C^{j_1j_2}_{JM}(m_1,m_2)$ are the Clebsch-Gordan coefficients.
The normalization constant $N$ is equal to $1/\sqrt{2}$ for nonequivalent electrons and to $1/2$ for equivalent electrons.

\par
Expansion of the one-electron wave function with certain momentum and polarization over the wave functions with certain total angular momentum and parity
(see
Eqs.\ \Br{eqn090514n01}, \Br{eqn110919n1})
and integration over the direction of the momentum of the incident electrons yield
\begin{eqnarray}
\sigma_{\ini\fin}
&=&\nonumber
\lim\limits_{N_{\oee_R}\to\infty}
\frac{\omega^2}{(2\pi)^2}
\left[
\frac{\oee}{p}
\frac{1}{4\pi}
\,
N_{\oee_R}^4
\,
\frac{\oee}{2\pi p N_{\oee_R}^2 S}
\left(\frac{(2\pi)^3}{\oee p}\right)^2
\right]
\\
&&\label{eqn110919n3}
\times
\sum\limits_{J,M,j_1\le j_2, l_1\le l_2, j_3\le j_4, l_3\le l_4}
A_{JM\oee_R j_ll_1 \oee_R j_2l_2 \oee_R j_3l_3 \oee_R j_4l_4}
\int d^2\hzhk\,
U_{JM\oee_R j_ll_1 \oee_R j_2l_2,\zhk\lambda s}
U^{*}_{JM\oee_R j_3l_4 \oee_R j_3l_4,\zhk\lambda s}
\,,
\end{eqnarray}
where coefficients $A_{JM\oee_R j_ll_1 \oee_R j_2l_2 \oee_R j_3l_3 \oee_R j_4l_4}$ are defined in
Appendix~\ref{appendixa}.
The last factor in the square brackets in
\Eq{eqn110919n3}
comes from the expansion of the one-electron wave function over the wave functions with certain momentum and polarization
(see
Appendix~\ref{appendixa}).
The matrix element of the recombination amplitude is calculated with two two-electron wave functions: ($JM\oee_R jl \oee_R j'l'$) for the initial state and $s$ for the final state.

%
\section{Calculation of the amplitude}
\label{electronx}
Following the notations employed in
\cite{andreev08pr},
we introduce the photon emission matrix elements
\begin{eqnarray}
A^{(k,\lambda)}_{ud}
&=&\label{eqn110929n1}
\int d\zhr\,
\bpsi_{u}(\zhr)
\gamma^{\mu}A^{(k,\lambda)}_{\mu}(\zhk)
\psi_{d}(\zhr)
\end{eqnarray}
and the one-photon exchange matrix elements
\begin{eqnarray}
I_{u_1u_2d_1d_2}(\Omega)
&=&\label{eqn110929n2}
\int d\zhr_1 d\zhr_2 \,
\bpsi_{u_1}(\zhr_1) \bpsi_{u_2}(\zhr_2)
\gamma^{\mu_1}_{1}\gamma^{\mu_2}_{2}
I_{\mu_1 \mu_2}(\Omega,r_{12})
\psi_{d_1}(\zhr_1) \psi_{d_2}(\zhr_2)
\,.
\end{eqnarray}
The indices $u_i$, $d_i$ designate one-electron Dirac states,
Dirac matrices $\gamma^{\mu_i}_{i}$ act on the one-electron functions $\psi_{d_i}(\zhr_i)$, respectively.
The photon wave function $A^{(k,\lambda)}$ is defined by
\Eq{photon-wavefunction-norm}.
Function $I_{\mu_1 \mu_2}(\Omega,r_{12})$ looks like
\begin{eqnarray}
I_{\mu_1 \mu_2}(\Omega,r_{12}) \label{ic}
&=&
\frac{\delta_{\mu_1 0} \delta_{\mu_2 0}}{r_{12}}
\\
&&\nonumber
-
\left(\frac{\delta_{\mu_1 \mu_2}}{r_{12}}\, e^{i\Omega r_{12}} +
\frac{\partial}{\partial x_1^{\mu_1}}
\frac{\partial}{\partial x_2^{\mu_2}} \frac{1}{r_{12}}\,
\frac{1-e^{i\Omega r_{12}}}{\Omega^2} \right)
\\
&&\label{it} \label{ib}
\times
(1-\delta_{\mu_1 0})(1-\delta_{\mu_2 0})
\,,
\end{eqnarray}
if Coulomb gauge is employed, or
\begin{eqnarray}
I_{\mu_1 \mu_2}(\Omega,r_{12}) 
&=&
\frac{g_{\mu_1 \mu_2}}{r_{12}}\, e^{i\Omega r_{12}}
\label{itfeynman} \label{ibfeynman}
\,,
\end{eqnarray}
if Feynman gauge is employed.
Tensor $g_{\mu_1 \mu_2}$ is the metric tensor; $\delta_{\mu_1 \mu_2}$ is the Kronecker delta; $r_{12}=|\zhr_1-\zhr_2|$.
Repeated indices imply summation.

\par
\begin{figure}
\centerline{\mbox{\epsfysize=100pt \epsffile{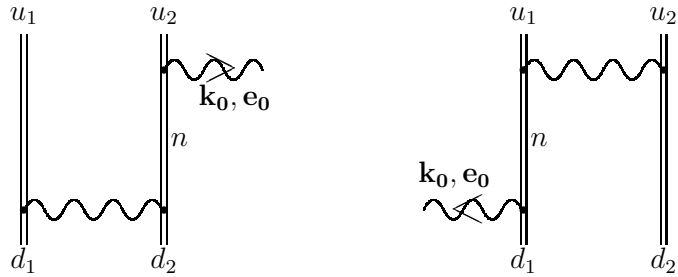}}}
\caption{\label{figuret02}
The Feynman graphs representing the first order interelectron interaction corrections to the process of electron recombination.
The internal wavy line denotes the exchange by the photon between two electrons.
The indices $d_1$, $d_2$ correspond to the initial one-electron states of a system; $u_1$, $u_2$ correspond to the final states.
the index $n$ corresponds to the intermediate one-electron states.
}
\end{figure}

The amplitudes presented in
\Eq{eqn110919n3}
are defined by Feynman graphs depicted in
Fig.\ \ref{figuret02}.
These amplitudes are proportional to the following expressions
\begin{eqnarray}
\xi_1
&=&\label{eqnxi1}
\sum\limits_{n}\xi_{1,n}
\,=\,
\sum\limits_{n}
\frac{A_{u_2 n}I_{u_1 n d_1 d_2}}{\dee_{u_1}+\dee_{n}-\dee_{d_1}-\dee_{d_2}}
\,,
\\
\xi_2
&=&\label{eqnxi2}
\sum\limits_{n}\xi_{2,n}
\,=\,
\sum\limits_{n}
\frac{I_{u_1 u_2 n d_2}A_{n d_1}}{\dee_{u_1}+\dee_{u_2}-\dee_{n}-\dee_{d_2}}
\,.
\end{eqnarray}
Here, $\xi_i$ corresponds the contributions of the left ($i=1$) and the right ($i=2$) graphs in
Fig.\ \ref{figuret02}, respectively.
The indices $d_1,d_2$ correspond to the continuum electrons of the initial state.
The indices $u_1,u_2$ correspond to the bound electrons of the final state.
The index $n$ may correspond to any state of the Dirac spectrum.

\par
If we consider an ion enclosed within a sphere of radius $R$, then for the states corresponding to the continuum we can write
(see
Eqs.\ \Br{eq080823n07}, \Br{eq080825n03})
\begin{eqnarray}
\psi_{\oee}(\zhr)
&\sim&\label{eqn110919n5}
\frac{1}{R^{1/2}}
\,,\phantom{123}
\Delta \oee
\sim
\frac{1}{R}
\,,
\end{eqnarray}
where $\Delta \oee$ is the distance between two closest energy levels.
The asymptotics ($R\to\infty$) of the matrix elements $A_{u d}$ and $I_{u_1 u_2 d_1 d_2}$ are investigated in
Appendix~\ref{appendixb}.
Here, we will investigate the behavior of the terms $\xi_i$ ($i=1,2$) with various values of the intermediate electron state $n$ when $R\to\infty$.

\par
If $n$ belongs to the discrete part of the spectrum, the terms $\xi_{i,n}$ contain two wave functions of electrons from the continuous part of the spectrum: $d_1$ and $d_2$ states.
Employing
Eqs.\  \Br{eqn111016n12}, \Br{eqn111016n3}
for $\xi_{1,n}$ and
Eqs.\ \Br{eqn111016n13}, \Br{eqn111016n10}
for $\xi_{2,n}$ we can write
\begin{eqnarray}
\xi_{i,n}
&\sim&\label{eqn110624n00}
\frac{1}{R}
\,,
\quad
i=1,2
\,.
\end{eqnarray}
The denominators in
Eqs.\ \Br{eqnxi1}, \Br{eqnxi2}
do not contain any smallness in this case.

\par
If $n$ belongs to the continuous part, then in general case
\begin{eqnarray}
\xi_{i,n}
&\sim&\label{eqn110624n01}
\frac{1}{R^{2}}
\,,
\quad
i=1,2
\,.
\end{eqnarray}
Formulas
\Br{eqn111016n10}, \Br{eqn111016n3}
and
 \Br{eqn111016n14}, \Br{eqn111016n11}
are used for $\xi_{1,n}$ and $\xi_{2,n}$, respectively.
The denominators in
Eqs.\ \Br{eqnxi1}, \Br{eqnxi2}
are supposed not to contain any smallness.
Consider now three special cases when
\Eq{eqn110624n01}
is violated.
The first case is when the energy of the intermediate electron state $n$ is equal to the energy of the incident electron ($\dee_n=\oee_e$): then, application of
Eqs.\ \Br{eqn111016n10}, \Br{eqn111016n9}
yields
\begin{eqnarray}
\xi_{1,n}
&\sim&\label{eqn110624n02}
\frac{\log(R)}{R^{2}}
\,.
\end{eqnarray}
The denominator in
\Eq{eqnxi1}
does not contain any smallness.
For description of the two other cases it is convenient to introduce a continuum electron $\tilde{e}$ with the energy
\begin{eqnarray}
\oee_{\tilde{e}}
&=&\label{eqn111013n1}
\oee_{e}+\oee_{e}-\oee_{1s}
\,.
\end{eqnarray}
The second special case is when the energy of the intermediate state $n$ is equal to the energy of the electron $\tilde{e}$ ($\dee_n=\oee_{\tilde{e}}$), then $\xi_{1,n}\to\infty$.
Here, the energy of the intermediate two-electron state ($\dee_{1s}+\dee_{n}$) is equal to the energy of the initial two-electron state ($\oee_{e}+\oee_{e}$).
The last case is when $\dee_n\approx \oee_{\tilde{e}}$.
If $\dee_n$ is the next state to $\oee_{\tilde{e}}$ (i.e., $\dee_n= \oee_{\tilde{e}}\pm \Delta\oee$), then
\begin{eqnarray}
\xi_{1,n}
&\sim&\label{eqn110624n03}
\frac{1}{R}
\,.
\end{eqnarray}
Here, formulas
\Br{eqn111016n10}, \Br{eqn111016n3}
are used for $A_{u_2n}$ and $I_{u_1nd_1d_2}$, respectively.
The denominator in
\Eq{eqnxi1}
is set to $\Delta\oee$ given by
\Eq{eq080823n07}.
Consider contribution of the states contained in the interval $(\oee_{\tilde{e}},\oee_{\tilde{e}}+\delta\oee]$, where $\delta\oee$ is a small finite value which does not depend on $R$.
The number ($K\approx\delta\oee/\Delta\oee$) of intervals $\Delta\oee$ composing the interval $\delta\oee$ is proportional to $R$.
Accordingly, we get
\begin{eqnarray}
\sum\limits_{\dee_n\in(\oee_{\tilde{e}},\oee_{\tilde{e}}+\delta\oee]}\xi_{1,n}
&\sim&\label{eqn110624n04}
\sum\limits^{K}_{k=1}
\frac{1}{kR}
\,\sim\,
\frac{\log(R)}{R}
\,.
\end{eqnarray}
Note, that the terms given by
Eqs.\ \Br{eqn110624n00}-\Br{eqn110624n02},\Br{eqn110624n03},\Br{eqn110624n04}
vanish (faster than $R^{-1/2}$ or faster than $N_{\oee_R}^{-1}$) in
\Eq{eqn110919n3}
when $N_{\oee_R}\to\infty$.
Accordingly, they do not contribute to the limit in
\Eq{eqn110919n3}.
The nonvanishing terms come from the second special case where $\dee_n=\oee_{\tilde{e}}$.
In this case the denominator in
\Eq{eqnxi1}
is equal to zero and the standard perturbation theory is not applicable.
However, the two-electron configuration $(e_R,e_R)_{J}$ representing the initial state and the two-electron configuration $(1s,\dee_n)_{J}$, where $J$ is the total angular momentum, can be considered as quasidegenerate ones.

\par
We have introduced the artificial electron state ($\tilde{e}$) by the condition: $\oee_{\tilde{e}}+\oee_{1s} = \oee_{e}+\oee_{e}$.
The subindex $R$ at $\tilde{e}_R$ indicates that the corresponding wave function is normalized to unity over the sphere of radius $R$.
The configuration $(1s,\tilde{e}_R)_{J}$ has the same energy as the configuration $(e_R,e_R)_{J}$.
The contribution of these configurations can be calculated within the framework of the LPA.
The configurations $(1s,\tilde{e}_R)_{J}$ and $(e_R,e_R)_{J}$ are considered as quasidegenerate ones.

\par
Within the LPA
\cite{andreev08pr}
we compose the matrix $V$
\begin{eqnarray}
V
&=&\label{eqd070723n02yyyz}
V^{(0)}+\Delta V
\,.
\end{eqnarray}
The matrix $V^{(0)}$ is a diagonal matrix and it includes the one-electron Dirac energies corresponding to a certain configuration.
The matrix $\Delta V$ includes the one-photon exchange corrections as well as other corrections which can be omitted here.
The matrix $V$ can be written in a block form
\begin{eqnarray}
V
&=&
\xmatrix{V_{11}}{\Delta V_{12}}{\Delta V_{21}}{V_{22}}
\,.
\end{eqnarray}
where the block $V_{11}$ contains matrix elements constructed on the configurations mixing with the reference state (the initial or final state).
The mixing configurations define the set $\setg$.
The block $V_{22}$ contains matrix elements calculated with all the other configurations, the blocks $\Delta V_{12}$, $\Delta V_{21}$ contains matrix elements constructed on one configuration from the set $\setg$ and one not included in the set $\setg$.

\par
Consider the set $\setg$ including only two configurations $(1s,\tilde{e}_R)_{J}$ and $(e_R,e_R)_{J}$ given by the two-electron wave functions of the noninteracting electrons in the \jj coupling scheme ($\Psi^{(0)}_{(1s,\tilde{e}_R)_{J}}$ and $\Psi^{(0)}_{(e_R,e_R)_{J}}$, respectively), where $J$ is the total angular momentum.
Then, the block matrix $V_{11}$ is $2\times2$ matrix
\begin{eqnarray}
V_{11}
&=&
\xmatrix{(V_{11})_{11}}{(\Delta V_{11})_{12}}{(\Delta V_{11})_{21}}{(V_{11})_{22}}\,.
\end{eqnarray}
Note, that the matrix $\Delta V$ is composed by one-photon exchange matrix elements which include continuum electron wave functions, therefore, they vanish when $R\to\infty$.
Accordingly, $(V_{11})_{11}=(V_{11})_{22}$ and $(\Delta V_{11})_{12}=(\Delta V_{11})_{21}\to 0$, when $R\to\infty$.
The eigenvectors of the matrix $V_{11}$ are
\begin{eqnarray}
{\zh b}_1
&\approx&\frac{1}{\sqrt{2}}\left({1}\atop{1}\right)
\,,\phantom{12}\,
{\zh b}_2
\,\approx
\,\frac{1}{\sqrt{2}}\left({-1}\atop{1}\right)
\,.
\end{eqnarray}
The eigenvectors ($\Phi$) of the operator $\hat{V}$ (represented by the matrix $V$) can be constructed with employment of the perturbation theory
\cite{andreev08pr}.
In the zeroth order of the perturbation theory, the eigenvectors ($\Phi_{(e_R,e_R)_{J}}$ and $\Phi_{(1s,\tilde{e}_R)_{J}}$) are combinations of the  two-electron wave functions of the noninteracting electron in the \jj coupling scheme ($\Psi^{(0)}_{(1s,\tilde{e}_R)_{J}}$ and $\Psi^{(0)}_{(e_R,e_R)_{J}}$)
\begin{eqnarray}
\Phi^{(0)}_{(e_R,e_R)_{J}}
&=&
({\zh b}_1)_2\Psi^{(0)}_{(e_R,e_R)_{J}}
+
\eta
({\zh b}_1)_1\Psi^{(0)}_{(1s,\tilde{e}_R)_{J}}
\\
\Phi^{(0)}_{(1s,\tilde{e}_R)_{J}}
&=&
\eta
({\zh b}_2)_2\Psi^{(0)}_{(e_R,e_R)_{J}}
+
({\zh b}_2)_1\Psi^{(0)}_{(1s,\tilde{e}_R)_{J}}
\,.
\end{eqnarray}
The factor $\eta=\pm 1$ defines which eigenvector corresponds to $(e_R,e_R)_{J}$ configuration.
It can be determined by the sign of the mixing element $(\Delta V_{11})_{12}$.

\par
The wave function $\Psi^{(0)}_{(e_R,e_R)_{J}}$ is proportional to $R^{-1}$ (or to $N_{\oee_R}^{-2}$), and therefore its contribution vanishes in
\Eq{eqn110919n3}
when $N_{\oee_R}\to\infty$.
The wave function $\Psi^{(0)}_{(1s,\tilde{e}_R)_{J}}$ is proportional to $R^{-1/2}$ (or to $N_{\oee_R}^{-1}$) because it contains only one continuum electron.
Accordingly, the contribution of the eigenvectors $\Phi_{(e_R,e_R)_{J}}$, $\Phi_{(1s,\tilde{e}_R)_{J}}$ is given by the contribution of the wave function $\Psi^{(0)}_{(1s,\tilde{e}_R)_{J}}$.
Note that the admixture of the $(1s,\tilde{e}_R)_{J}$ configuration to the $(e_R,e_R)_{J}$ configuration leads to a growth of the amplitude with factor $N_{\oee_R}$.
This growth is compensated by the additional factor $1/V$ in the transition probability for the RDEC process (see
Eqs.\ \Br{eqn090507n01} and \Br{eqn110919n2}).

\par
We can conclude that, if we take into account the interelectron interaction (the one-photon exchange corrections in all orders of the perturbation theory) between the $1s$, $e_R$ and $\tilde{e}_R$ electron states, the amplitudes of the processes
\begin{eqnarray}
e+e\to (1s1s)+\gamma(\omega)
\phantom{12}\mbox{and}\phantom{12}
1s+\tilde{e}\to (1s1s)+\gamma(\omega)
&&\nonumber
\end{eqnarray}
are connected as
\begin{eqnarray}
U[e+e]
&=&
\eta
U[1s+\tilde{e}]
\,.
\end{eqnarray}
This equation is valid up to the terms disappearing when $N_{\oee_R}\to\infty$.

%
\section{Numerical methods}
\label{methods}
The Dirac spectrum is constructed in our work in terms of B-splines
\cite{johnson88p307,shabaev04}.
The ion is placed into a spherical box with the radius $R_\txt{B}=70/(\alpha Z)$ (in the relativistic units), where $Z$ is the nuclear charge and $\alpha$ is the fine-structure constant.
The B-splines used in our calculations are of the order $8$ and we employed the grid with $60$ nonzero knots.
The generated electron spectrum is discrete and finite.

\par
The eigenvector and the corresponding eigenvalue (\ie, energy), which is the closest to the energy of the continuum electron ($\oee_{\tilde{e}}$), is replaced by the wave function of the electron confined within the spherical box of radius $R$
(\Eq{eq110403n04}),
and by the energy $\oee_{\tilde{e}}$, respectively.
The electron states of the generated spectrum, which are close to the substituted electron state ($e_{n}$), are designated as $e_{n-1}$ and $e_{n+1}$.
Extension of the number of the knots and the size of the box reduces the effect of the substitution of the continuum electron state ($\oee_{\tilde{e}}$) by the $e_{n}$ state.
After some variations and tests with different conditions (the number of knots and the number of states when we sum over Dirac spectrum) we choose the number of knots and the size of the box to be large enough not to influence the accuracy of the computing.

\par
In the present paper we consider the RDEC to the ground state $(1s1s)$ and to the low lying single excited states (the $KL$-shell).
The contribution of the $KL$-shell states is determined by contribution of the $(1s2s)_0$ configuration.
The contribution of the other states does not exceed $1\%$ of the total cross-section.
We also consider only the main channel of the RDEC: capture with emission of the electric photons with $J=1$.

\par
The LPA
\cite{andreev08pr,andreev09p042514}
is employed for calculation of the amplitude of the RDEC.
In the framework of the LPA we fix a set of electron states and construct a set $\setg$ of all possible two-electron configurations in the \jj coupling scheme built on this set of electron states.
We introduce $ns$, $np$-electron states as the electron states in the B-spline approximation which energy is closest to the energy of the continuum electron state $\oee_{\tilde{e}}$.
The electrons included into the set $\setg$ are $1s$, $2s$, $2p$, $ns$, $np$, $(n\pm1)s$, $(n\pm1)p$-electron states.
The $(n\pm1)s$, $(n\pm1)p$ are electron states in the B-spline approximation next to the $ns$, $np$-electron states.

\par
In the numerical calculations we construct the matrix $V$ in a special way to include the contribution of the mixing configurations
\cite{andreev08pr,andreev09p042514}.
The matrix $V$ is calculated with application of the QED perturbation theory
\begin{eqnarray}
V(\omega)
&=&\label{eqd070723n02yyy}
V^{(0)}+\Delta V
\,=\,
V^{(0)}+ V^{(1)}(\omega)
+\ldots
\,.
\end{eqnarray}
In physical sense the $\omega$ is the frequency of the scattered photon and the matrix $V$ depends on the value of the $\omega$.
The matrix $V^{(0)}$ is composed with one-electron Dirac energies corresponding to a certain configuration.
The matrix $V^{(1)}(\omega)$ includes the first order QED corrections, such as self-energy (SE) and vacuum polarization (VP) corrections and one-photon exchange corrections.
In our work the matrix $V(\omega)$ contains only $V^{(0)}$ and $V^{(1)}(\omega)$, the last one includes only one-photon exchange corrections:
\begin{eqnarray}
\Delta V^\txt{1ph}_{u_1u_2d_1d_2}
&=&
I(|\dee_{u_2}-\dee_{d_2}|)_{u_1u_2d_1d_2}
\end{eqnarray}
(see
\Eq{eqn110929n2}).
Within the framework of the LPA the contribution of the one-photon exchange correction is taken into account to all orders of the perturbation theory for the configurations from the set $\setg$. 

\par
The amplitude of the transition from the initial state $\Ini$ to the final state $\Fin$ with emission of one photon with the frequency $\omega_{0}$ can be written as
\cite{andreev08pr}
\begin{eqnarray}
U_{\Ini\to \Fin}
&=&
\left(\Xi(\omega_{0})\right)_{\Phi_{\Fin}\Phi_{\Ini}}
\,,
\end{eqnarray}
where $\Xi(\omega_{0})$ is operator of emission of the photon, $\Phi_{\Ini}$ and $\Phi_{\Fin}$ are the eigenvectors of the matrix $V(\omega)$ corresponding to the $\Ini$ and $\Fin$ states, respectively.
The operator $\Xi(\omega_{0})$ is evaluated with employment of the QED perturbation theory (see
\cite{andreev08pr,andreev09p042514}).
In zero order approximation this operator coincides with the photon emission operator ($A^{(k_0,\lambda_0) *}$).
In this work we consider only the one-photon exchange corrections to the operator $\Xi$.
According  to
\cite{andreev09p042514},
it reads
\begin{eqnarray}
\Xi
&=&\label{x050520x25}
\Xi^{(0)} + \Xi^{(1)} + e\Ordnung
(\alpha^2)
\,.
\end{eqnarray}
The zero-order matrix element is
\begin{eqnarray}
{ \Xi}^{(0)}_{u_1 u_2 d_1 d_2}
&=&\label{x050520x3}
2e
A^{(k_0,\lambda_0) *}_{u_1 d_1} \delta_{u_2 d_2}
\,,
\end{eqnarray}
where $A^{(k_0,\lambda_0)*}_{n_1 n_2}$ are the matrix elements of the emission operator which includes the photon wave function
\Eq{photon-wavefunction-norm}.

\par
It is convenient to write the matrix $V$ in a block form
\begin{eqnarray}
V
&=&
\xmatrix{V_{11}}{V_{12}}{V_{21}}{V_{22}}
=
\xmatrix{V^{(0)}_{11}+\Delta V_{11}}
        {\Delta V_{12}}{\Delta V_{21}}
        {V^{(0)}_{22}+\Delta V_{22}}
\,,
\end{eqnarray}
where the block $V_{11}$ is composed entirely with the states from the set $\setg$ and the block $V_{22}$ does not contain states from the set $\setg$.
The blocks $V_{12}$ and $V_{21}$ are constructed with one configuration from the set $\setg$ and with another one not included in the set $\setg$.

\par
The matrix $V_{11}$ can be diagonalized numerically (non-perturbatively)
\begin{eqnarray}
V^\txt{diag}_{11}
&=&\label{eqd070723n01}
B^{t}V_{11} B
\,,
\end{eqnarray}
where $B$ is an orthogonal matrix, $B^t$ is the transposed matrix.
Since in general $V$ is a complex symmetric matrix, the matrix $B$ is a complex orthogonal matrix
\begin{eqnarray}
B^{t}B
&=&\label{eqd070206n01}
I
\,.
\end{eqnarray}
where $I$ is a unit matrix ($I_{ij}=\delta_{ij}$) of the appropriate
dimension.

\par
The eigenvector of the matrix $V$ can be written as
\cite{andreev08pr}.
\begin{eqnarray}
\Phi_{n_{\setg}}
&=&\label{x050525x06}
\sum\limits_{k_{\setg}\in\setg}
B_{k_{\setg} n_{\setg}} \Psi^{(0)}_{k_{\setg}}
+ \sum\limits_{{k\notin\setg}\atop {l_{\setg}\in\setg} }
(\Delta V_{21})_{k l_{\setg}}
\frac{B_{l_{\setg}n_{\setg}}}
{E_{n_{\setg}}-E^{(0)}_{k}}
\Psi^{(0)}_{k}
+
\ldots
\,,
\end{eqnarray}
where $E_{n_{\setg}}$ are the eigenvalues of the matrix $V_{11}$ and
$E^{(0)}_{k}$ are the sums of the Dirac energies.
The functions $\Psi^{(0)}$ are the two-electron functions in the \jj coupling scheme.
The indices $k_{\setg}$, $l_{\setg}$ run over configurations from the set $\setg$, while the index $k$ runs over configurations not included in the set $\setg$, \ie, over all the other two-electron configurations.
The first term in the right-hand side of the expression
\Eq{x050525x06}
can be considered as the zero order of the applied perturbation theory, the second term corresponds to the first order.

\par
The cross-section is given by
\Eq{eqn110919n3},
where the amplitude $U$ enters as its squared absolute value.
Employing
\Eq{x050525x06},
the amplitude $U$ can be presented as $U=U^{(0)}+U^{(1)}+\ldots$.
Accordingly, the squared absolute value of $U$ can be written as $|U|^2=|U^{(0)}|^2+2\real\{U^{(0)}U^{(1)}\}+|U^{(1)}|^2$.
The last term corresponds to the second order corrections and, in principle, can be omitted.
Still we prefer to keep it.
We consider the contribution of this term as an estimate of magnitude of the higher order corrections.

\par
To compare our results with experiment we use in our calculation the model of two electrons which are moving along the same direction with equal momentum.

\par
The calculation was performed with different gauges for the exchange-photon (the Coulomb and Feynman gauges) and emitted photons (the transverse and nontransverse gauges
\cite{andreev08pr}).
A small deviation of the gauge invariance took place.
It is explained by the fact that the set of Feynman graphs that we take into account is not gauge invariant beyond the lowest QED PT order.
The magnitude of the deviation is determined by the magnitude of the higher order corrections.

%
%
\section{Results and discussion}
\label{results}
In our work we calculate the cross-section for double electron capture by a bare nucleus followed by emission of the photon.
We calculate the cross-section for three different experiments and present our results in
Tables~\ref{tabO1}-\ref{tabU1}.
We consider the scenario when two electrons are going along one line and have the same value of the momentum $\zhp$.
In our model the incident electrons are considered as Dirac continuum electrons.
According to
Eqs.\ \Br{eqn090507n01}, \Br{eqn111016n21}
the cross-section of the RDEC process depends on the volume of reaction.
In the RDEC experiments the captured electrons are initially located on an atom.
We consider two approximation for the experiments: 1) we suppose that the electrons are distributed homogeneously in the atom ($\sigma^\txt{RDEC,A}$); 2) we neglect all the electrons of the atom except the $K$-shell electrons and suppose that the electrons are distributed homogeneously within the sphere of the $K$-shell radius of the atom ($\sigma^\txt{RDEC,K}$).

\par
In
Tables~\ref{tabradiusC}, \ref{tabradiusAr}
we present the radii and areas of the cross-sections of the target atoms (see
Section~\ref{electron})
used for the calculation of the RDEC cross-section.
The radius $R^\txt{A}_\txt{T}$ is the radius of the target atom.
The area of the cross-section of the reaction volume for one electron ($S^\txt{A}$) is calculated as $S^\txt{A}=\pi (R^\txt{A}_\txt{T})^2/Z_\txt{T}$, where $Z_\txt{T}$ is the charge of the nucleus of the target atom.
The area $S^\txt{A}$ is employed in the case of the first approximation.
The radius $R^\txt{K}_\txt{T}$ is the radius of the $K$-shell of the target atom.
In the case of the second approximation the area of the cross-section of the reaction volume for one electron ($S^\txt{K}$) is set to $S^\txt{K}=\pi (R^\txt{K}_\txt{T})^2/2$.
In the case of the second approximation ($\sigma^\txt{RDEC,K}$) the expressions for the RDEC probability and cross-section
Eqs.\ \Br{eqn090507n01}, \Br{eqn111016n21}, \Br{eqn110919n3}
get an additional factor $(R^\txt{K}_\txt{T}/R^\txt{A}_\txt{T})^3$ (ratio between the volume of the $K$-shell and the volume of the whole target atom).
Accordingly, the values of $\sigma^\txt{RDEC,A}$ and $\sigma^\txt{RDEC,K}$ differs by a factor
\begin{eqnarray}
&&
\frac{S^\txt{A}}{S^\txt{K}}
\left(\frac{R^\txt{K}_\txt{T}}{R^\txt{A}_\txt{T}}\right)^3
\,.
\end{eqnarray}

\par
First, we consider the case of the RDEC process when the ion of oxygen with the energy $38$ MeV is hitting a carbon foil.
The density of the carbon foil is of the order of $10^{17}$ particles/cm$^{2}$.
The results of our calculation for the RDEC process O${}^{8+}+$C we present in the Tables~\ref{tabO1}, \ref{tabO2}, \ref{tabO3}.
In
Table~\ref{tabO1}
we present the data for the RDEC process to the ground state.
The first column presents the experimental value of the cross-section, the second gives the results of the nonrelativistic calculation
\cite{mikhailov04p350}.
The last two columns present our results.
Data for the cross-section of the RDEC process to the $(1s2s)$ state are given in
Table~\ref{tabO2}.
The RDEC to the $(1s2s)$ determine actually the contribution of the RDEC to the all single excited states of the $KL$-shell since the contribution of the states higher than $(1s2s)$ are quite small.
The total cross-section of the RDEC for oxygen ($\sigma^\txt{RDEC}=\sigma_{(1s1s)}^\txt{RDEC}+\sigma_{(1s2s)}^\txt{RDEC}$)
is presented in
Table~\ref{tabO3}.
The experiment of the RDEC in collisions of O${}^{8+}$ ions with carbon was reported in
\cite{simon10},
though the most detailed description of the experiment is presented in
\cite{simon10phd}.
We note that the the experimental data
\cite{simon10phd}
for the separate contributions of the $(1s1s)$ and $(1s2s)$ configurations were defined with the use of the calculations
\cite{mikhailov04p350,nefiodov05}.
It explains good agreement between the ratio of these contributions defined in
\cite{simon10phd}
and in
\cite{mikhailov04p350,nefiodov05}.

\par
The RDEC experiment for argon was performed with $11.4$~MeV/u Ar${}^{18+}$ ions hitting a carbon foil at the UNILAC in GSI in 1994
\cite{warczak95}.
The thickness of the target was 4-10 $\mu \mbox{g/cm}^{2}$.
In this experiment the probability of the RDEC process is very small.
Experimental data give only the upper limit of the value of the cross-section of the process: about $5.2$~mb.
The results of the calculations for RDEC process Ar${}^{18+}+$C are presented in the
Table~\ref{tabAr1}.

\par
The RDEC experiment for uranium was performed at the heavy ion storage ring (ESR) in GSI
\cite{bednarz03}.
Bare U${}^{92+}$ ions with the energy of $297$~MeV/u have been used in collisions with gas of Ar atoms.
The density of the gaseous Ar-target was $5\times 10^{12}$ particles$/\mbox{cm}^{2}$.
The data obtained in this experiment provide only the upper limit of the cross-section: less than $10$~mb.
The upper limit of RDEC process was estimated to be significantly lower than measured previously
\cite{bednarz01}.
The results of the calculations for RDEC process U${}^{92+}+$Ar are presented at the
Table~\ref{tabU1}.

\par
Comparison of the results for the RDEC cross-section reveals disagreement between the experimental and theoretical data.
The disagreement between the experimental data and our results can be explained by the model, which we employed for description of the target used in the experiments
\cite{simon10,warczak95,bednarz01}.
This model is rather rough.
In the experiments the captured electrons are initially the bound electrons of atoms of either carbon foil
\cite{simon10,warczak95}
or argon gas
\cite{bednarz01}.
In our model the incident electrons are considered as Dirac continuum electrons.
In particular, we do not take into account the bound energy of the target electrons.
We also suppose that the electron density in the target atoms is homogeneous.
The disregard of the bound energy should exaggerates the results.
The assumption that the electrons in the target atom are distributed homogeneously should also change the results, however, it is difficult to estimate its influence.

\par
The disagreement between the theoretical results obtained in
\cite{mikhailov04p350}
and our results is not clear.
In principle, the models employed for description of the target electrons are rather similar.
The work
\cite{mikhailov04p350}
presents nonrelativistc calculation, while our one is fully relativistic.
However the relativistic effects can not explain the present disagreement.
We note that the volume of reaction (where the incident electrons interact with the bare nucleus) is defined different in these calculations.
In our calculation the reaction volume is a cylinder (see
\Fig{fig-cylinder}),
while in work
\cite{mikhailov04p350}
the reaction volume is a sphere of radius $R_\txt{T}$, where $R_\txt{T}$ is the radius of the target atom in its rest frame.
We also note that the ratios between the cross-sections of the RDEC to the ground state $(1s1s)$ and to the $KL$-shell $(1s2s)$ (see
Tables~\ref{tabO1}, \ref{tabO2})
obtained in these calculations are different.
According to our calculation the total RDEC cross-section is determined by the RDEC to the ground state, while the results of
\cite{mikhailov04p350,nefiodov05}
predict that the main contribution to the RDEC is given by the capture to the excited states.

\par
The model of quasifree electrons employed in this calculation is rather rough for description of the experiments
\cite{simon10,warczak95},
where relatively light bare nuclear ($Z=8,18$) move trough carbon foil ($Z=6$).
Experiments where the target atoms are much lighter than the bare nucleus
(e.g., experiment
\cite{bednarz01})
are preferable.
This model would be also good for description of experiments with electron beams.
The small relative velocity between the bare nucleus and electrons and the presence of the magnetic field could enlarge the cross-section
\cite{shi01}.

%
%
\begin{acknowledgments}
The helpful discussions with A.I. Mikhailov and A.V. Nefiodov are gratefully acknowladged.
This work was supported by the RFBR grant 11-02-00168-a.
\end{acknowledgments}

%
%
\appendix
\section{Asymptotics of matrix elements}
\label{appendixb}
The asymptotics of the Dirac wave functions for electrons in the continuum is given by
Eqs.\ \Br{eq080824n01}-\Br{eq080823n06}.
These wave functions are normalized to the Dirac delta function with respect to the energy
(\Eq{eqn110622}).
In case of the normalization of the wave functions of the continuum electrons to unity over a sphere of radius $R$ the normalization integral reads
\begin{eqnarray}
(N_{\oee_R})^2
&=&
\int\limits_{r\le R} d\zhr\, \psi_{\oee jlm}^{+}(\zhr)\psi_{\oee jlm}(\zhr)
\,.
\end{eqnarray}
Employing
Eqs.\ \Br{eq080824n01}-\Br{eq080823n06}
and performing integration over the angular variables ($\zhnu_r$) we get
\begin{eqnarray}
(N_{\oee_R})^2
&=&\label{eqn111016n6}
\int\limits^R_0 dr\, (|g_{\oee}(r)|^2 + |f_{\oee}(r)|^2)
\\
&=&\label{eqn111016n7}
\int\limits^R_0 dr\,
\left(
\frac{\oee+m}{\pi p} \cos^2(pr+\phi_{g}(r))
+
\frac{\oee-m}{\pi p} \sin^2(pr+\phi_{f}(r))
\right)
\\
&=&\label{eqn111016n8}
\int\limits^R_0 dr\,
\left(
\frac{\oee}{\pi p}
+
\frac{\oee+m}{2\pi p} \cos(2pr+2\phi_{g}(r))
+
\frac{\oee-m}{2\pi p} \sin(2pr+2\phi_{f}(r))
\right)
\,.
\end{eqnarray}
The first term of the integrand does not depend on $R$, its contribution is proportional to $R$.
The last two terms contain the sine and cosine functions, the absolute value of their contribution does not exceed a value not dependent on $R$.
So, we can write
\begin{eqnarray}
(N_{\oee_R})^2
&=&\label{eqn111016n20}
\frac{\oee R}{\pi p}
(1 + \Ordnung(R^{-1}))
\,.
\end{eqnarray}
Accordingly, the wave function of the continuum electron normalized to the Dirac delta function and the one normalized to unity over the sphere of radius $R$ are connected by
Eqs.\ \Br{eq110403n04}, \Br{eq110403n02}.

\par
Consider the asymptotics ($R\to\infty$) of the one-photon exchange matrix elements ($I_{u_1u_2d_1d_2}$) given by
\Eq{eqn110929n2}.
For this purpose we can restrict ourselves by the Coulomb part of the function $I_{\mu_1 \mu_2}(\Omega,r_{12})$, which looks like
\begin{eqnarray}
I_{\mu_1 \mu_2}(\Omega,r_{12})
&=&\label{icoulomb}
\frac{\delta_{\mu_1 0} \delta_{\mu_2 0}}{r_{12}}
\,.
\end{eqnarray}
It is convenient to employ the decomposition
\begin{eqnarray}
\frac{1}{r_{12}}
&=&\label{eqn111016n2}
\sum\limits^{\infty}_{k=0}
\frac{r_{<}^{k}}{r_{>}^{k+1}}
P_k(\zhnu_{r_1}\zhnu_{r_2})
\,,
\end{eqnarray}
where $r_{<}=\min(r_1,r_2)$,  $r_{>}=\max(r_1,r_2)$, $P_k(x)$ is the Legendre polynomial, $\zhnu_{r_1}\zhnu_{r_2}$ is the scalar product of vectors $\zhnu_{r_i}=\zhr_i/r_i$ ($i=1,2$).
For investigation of the asymptotics we can also retain only the term with $k=0$ in the decomposition
\Eq{eqn111016n2}.
So, we get
\begin{eqnarray}
I_{u_1u_2d_1d_2}
&=&\label{eqn111016n1}
\int\limits_{r_1\le R} d\zhr_1 \int\limits_{r_2\le R}d\zhr_2 \,
\psi^{+}_{u_1}(\zhr_1) \psi^{+}_{u_2}(\zhr_2)
\frac{1}{r_{>}}
\psi_{d_1}(\zhr_1) \psi_{d_2}(\zhr_2)
\,.
\end{eqnarray}
We assume that all the electron wave functions are normalized to unity over the sphere of radius $R$.
Accordingly, every continuum electron wave function has got a normalization factor $1/N_{\oee_R}$ and it is proportional to $1/\sqrt{R}$ (see
\Eq{eq080825n03}).
We will investigate the asymptotics ($R\to\infty$) of the matrix element
\Eq{eqn111016n1}
for various electron states $n$.

\par
At first we will suppose that the electron states $u_1$, $u_2$ correspond to bound electrons (e.g., $1s$-electron state), the electron state $d_1$ (we will designate it as $n$) can be any state of the Dirac spectrum, the electron state $d_2$ describes a continuum electron.
If the electron state $n$ corresponds to a bound electron, then there is only one continuum electron ($d_2$) containing factor $1/\sqrt{R}$ in the matrix element.
Accordingly, we get
\begin{eqnarray}
I_{u_1u_2nd_2}
&\sim&\label{eqn111016n13}
\frac{1}{R^{1/2}}
\,.
\end{eqnarray}
If the electron state $n$ corresponds to a continuum electron, then there are two continuum electron ($n$ and $d_2$) containing factor $1/\sqrt{R}$ in the matrix element
\begin{eqnarray}
I_{u_1u_2nd_2}
&\sim&\label{eqn111016n14}
\frac{1}{R}
\,.
\end{eqnarray}

\par
Now, we will suppose that the electron states $d_1$, $d_2$ correspond to continuum electrons with the same energies equal to $\oee$, the electron state $u_1$ corresponds to a bound electron (e.g., $1s$-electron state), the electron state $u_2$ (we will designate it as $n$) can be any state of the Dirac spectrum.
If the electron state $n$ corresponds to a bound electron, then there are only two continuum electrons ($d_1$ and $d_2$) containing factor $1/\sqrt{R}$ in the matrix element.
Accordingly, we get
\begin{eqnarray}
I_{u_1nd_1d_2}
&\sim&\label{eqn111016n3}
\frac{1}{R}
\,.
\end{eqnarray}
If the electron state $n$ corresponds to a continuum electron (different from the electron $d_2$), then there are three continuum electrons ($d_1$, $d_2$ and $n$) in the matrix element.
It yields
\begin{eqnarray}
I_{u_1nd_1d_2}
&\sim&\label{eqn111016n4}
\frac{1}{R^{3/2}}
\,.
\end{eqnarray}
We have to select a special case when the electron state $n$ coincides with electron $d_2$.
The matrix element $I_{u_1nd_1d_2}$ can be written as
\begin{eqnarray}
I_{u_1nd_1d_2}
&=&\nonumber
\int\limits_{r_1\le R} d\zhr_1\,
\psi^{+}_{u_1}(\zhr_1)\psi_{d_1}(\zhr_1)
\\
&&\label{eqn111016n5}
\times
\left[\frac{1}{r_1}
\int\limits_{r_2\le r_1} d\zhr_2 \,
\psi^{+}_{n}(\zhr_2) \psi_{d_2}(\zhr_2)
+
\int\limits_{r_1\le r_2\le R} d\zhr_2 \,
\psi^{+}_{n}(\zhr_2) \frac{1}{r_{2}} \psi_{d_2}(\zhr_2)
\right]
\,.
\end{eqnarray}
If the wave function $\psi_{n}$ is equal to $\psi_{d_2}$, then the last term in the square brackets is
\begin{eqnarray}
\int\limits_{r_1\le r_2\le R} d\zhr_2 \,
\psi^{+}_{n}(\zhr_2) \frac{1}{r_{2}} \psi_{d_2}(\zhr_2)
&=&
\int\limits_{r_1\le r_2\le R} d\zhr_2 \,
\psi^{+}_{d_2}(\zhr_2) \frac{1}{r_{2}} \psi_{d_2}(\zhr_2)
\\
&=&
N_{d_2}^{-2}
\int\limits^{R}_{r_1} dr_2 \,
\frac{1}{r_{2}}
\left(|g_{d_2}(r_2)|^2+ |f_{d_2}(r_2)|^2\right)
\,,
\end{eqnarray}
where the asymptotics of the $g_{d_2}$ and $f_{d_2}$ functions are given by
Eqs.\ \Br{eq080823n05}, \Br{eq080823n06}.
Employing
Eqs.\ \Br{eqn111016n6}-\Br{eqn111016n8}
we get
\begin{eqnarray}
\int\limits_{r_1\le r_2\le R} d\zhr_2 \,
\psi^{+}_{n}(\zhr_2) \frac{1}{r_{2}} \psi_{d_2}(\zhr_2)
&=&
\frac{1}{R}
(\log(R) + \Ordnung(R^{0}))
\,,
\end{eqnarray}
where the logarithmic term is given by the first term of the integrand in
\Eq{eqn111016n8}.
The first term in the square brackets in
\Eq{eqn111016n5}
is proportional to $1/R$.
Accordingly, for the case when the electron state $n$ coincides with the electron $d_2$ the matrix element $I_{u_1nd_1d_2}$ reads
\begin{eqnarray}
I_{u_1nd_1d_2}
&\sim&\label{eqn111016n9}
\frac{\log(R)}{R^{3/2}}
\,.
\end{eqnarray}

\par
Consider also the asymptotics of the photon emission matrix elements $A_{ud}$ given by
\Eq{eqn110929n1}.
We assume that all the electron wave functions are normalized to unity over the sphere of radius $R$.
If the both electrons ($d$ and $u$) describe the bound electrons then the matrix element $A_{ud}$ does not depend on $R$
\begin{eqnarray}
A_{ud}
&\sim&\label{eqn111016n12}
R^{0}
\,.
\end{eqnarray}
If one of the electrons ($d$ or $u$) corresponds to the bound electrons and the other one to the continuum electron then
\begin{eqnarray}
A_{ud}
&\sim&\label{eqn111016n10}
\frac{1}{R^{1/2}}
\,.
\end{eqnarray}
If the both electrons describe the continuum electrons, then
\begin{eqnarray}
A_{ud}
&\sim&\label{eqn111016n11}
\frac{1}{R}
\,.
\end{eqnarray}

%
%
\section{Angular integration}
\label{appendixa}
Integration over the direction of the incident electrons momentum in
\Eq{eqn090514n13}
can be performed analytically.
Only the two-electron wave function of the initial state depends on the direction of the momentum.
We consider the following integral
(see
Eqs.\ \Br{eqn090514n13}, \Br{eqn110919n4})
\begin{eqnarray}
I
&=&
\int d\zhnu_p\,
\frac{1}{\sqrt{2}}
\det\{\psi_{\zhp\mu_{p1}=1/2}(\zhr_1)\psi_{\zhp\um{\mu}_{p1}=-1/2}(\zhr_2)\}
\frac{1}{\sqrt{2}}
\det\{\psi^{*}_{\zhp\mu_{p3}=1/2}(\zhr_3)\psi^{*}_{\zhp\um{\mu}_{p3}=-1/2}(\zhr_4)\}
\\
&=&
\frac{1}{2}
\sum\limits_{\mu_{p1}\mu_{p3}}\int d\zhnu_p\,
(-1)^{-1/2+\mu_{p1}} (-1)^{-1/2+\mu_{p3}}
\psi_{\zhp\mu_{p1}}(\zhr_1)
\psi_{\zhp\um{\mu}_{p1}}(\zhr_2)
\psi^{*}_{\zhp\mu_{p3}}(\zhr_3)
\psi^{*}_{\zhp\um{\mu}_{p3}}(\zhr_4)
\end{eqnarray}
The one-electron wave functions with certain momentum and polarization can be expanded in series over the wave functions with certain total angular momentum and parity.
Employing
\Eq{eqn090514n01}
we get
\begin{eqnarray}
I
&=&\nonumber
\frac{1}{2}
\sum\limits_{\mu_{p1}\mu_{p3}}\int d\zhnu_p\,
\int d\dee_1 d\dee_2 d\dee_3 d\dee_4
\sum\limits_{j_1j_2j_3j_4 l_1l_2l_3l_4 m_1m_2m_3m_4 }
(-1)^{-1/2+\mu_{p1}} (-1)^{-1/2+\mu_{p3}}
\\
&&\nonumber
\times
a_{\zhp \mu_{p1},\dee_1 j_1l_1m_1}
a_{\zhp \um{\mu}_{p1},\dee_2 j_2l_2m_2}
a^{*}_{\zhp \mu_{p3},\dee_3 j_3l_3m_3}
a^{*}_{\zhp \um{\mu}_{p3},\dee_4 j_4l_4m_4}
\\
&&
\times
\psi_{\dee_1j_1l_1m_1}(\zhr_1)
\psi_{\dee_2j_2l_2m_2}(\zhr_2)
\psi^{*}_{\dee_3j_3l_3m_3}(\zhr_3)
\psi^{*}_{\dee_4j_4l_4m_4}(\zhr_4)
\end{eqnarray}
The coefficients $a_{\zhp \mu_{p},\dee j l m}$ can be written as
Eqs.\ \Br{eqn110919n1}, \Br{eqn110919n6}.
The integration over energies yields
\begin{eqnarray}
I
&=&\nonumber
\frac{1}{2}
\left(\frac{(2\pi)^3}{p\oee}\right)^2
\sum\limits_{\mu_{p1}\mu_{p3}}\int d\zhnu_p\,
\sum\limits_{j_1j_2j_3j_4 l_1l_2l_3l_4 m_1m_2m_3m_4 }
(-1)^{-1/2+\mu_{p1}} (-1)^{-1/2+\mu_{p3}}
\\
&&\nonumber
\times
e^{i\varphi_{\oee j_1l_1}}
\alpha_{\zhnu_p \mu_{p1},j_1l_1m_1}
e^{i\varphi_{\oee j_2l_2}}
\alpha_{\zhnu_p \um{\mu}_{p1},j_2l_2m_2}
e^{-i\varphi_{\oee j_3l_3}}
\alpha^{*}_{\zhnu_p \mu_{p3},j_3l_3m_3}
e^{-i\varphi_{\oee j_4l_4}}
\alpha^{*}_{\zhnu_p \um{\mu}_{p3},j_4l_4m_4}
\\
&&\label{eqn110919n7}
\times
\psi_{\oee j_1l_1m_1}(\zhr_1)
\psi_{\oee j_2l_2m_2}(\zhr_2)
\psi^{*}_{\oee j_3l_3m_3}(\zhr_3)
\psi^{*}_{\oee j_4l_4m_4}(\zhr_4)
\,.
\end{eqnarray}

\par
Consider separately the following integral
\begin{eqnarray}
I_1
&=&\nonumber
\sum\limits_{\mu_{p1}\mu_{p3}}\int d\zhnu_p\,
(-1)^{-1/2+\mu_{p1}} (-1)^{-1/2+\mu_{p3}}
\\
&&\label{eqn110919n8}
\times
\alpha_{\zhnu_p \mu_{p1},j_1l_1m_1}
\alpha_{\zhnu_p \um{\mu}_{p1},j_2l_2m_2}
\alpha^{*}_{\zhnu_p \mu_{p3},j_3l_3m_3}
\alpha^{*}_{\zhnu_p \um{\mu}_{p3},j_4l_4m_4}
\,.
\end{eqnarray}
Employing
\Eq{eqn110919n6}
yields
\begin{eqnarray}
I_1
&=&\nonumber
\sum\limits_{\mu_{p1}\mu_{p3}}\int d\zhnu_p\,
\!\!\!\!\!\!\!\!\!
\sum\limits_{m_{l1}m_{l2}m_{l3}m_{l4}\mu_1\mu_2\mu_3\mu_4}
\!\!\!\!\!\!\!\!\!
C^{l_1\frac{1}{2}}_{j_1m_1}(m_{l1},\mu_1)
C^{l_2\frac{1}{2}}_{j_2m_2}(m_{l2},\mu_2)
C^{l_3\frac{1}{2}}_{j_3m_3}(m_{l3},\mu_3)
C^{l_4\frac{1}{2}}_{j_4m_4}(m_{l4},\mu_4)
\\
&&\nonumber
\times
Y^{*}_{l_1m_{l1}}(\zhnu_p) Y^{*}_{l_2m_{l2}}(\zhnu_p)
Y_{l_3m_{l3}}(\zhnu_p) Y_{l_4m_{l4}}(\zhnu_p)
\\
&&\nonumber
\times
(-1)^{1+\mu_{p1}+\mu_{p3}}
\\
&&
\times
[\eta^{+}(\mu_1)v^{\mu_{p1}}(\zhnu_p)]
[\eta^{+}(\mu_2)v^{\um{\mu}_{p1}}(\zhnu_p)]
[\eta^{+}(\mu_3)v^{\mu_{p2}}(\zhnu_p)]^{*}
[\eta^{+}(\mu_4)v^{\um{\mu}_{p2}}(\zhnu_p)]^{*}
\,.
\end{eqnarray}
After integration over the direction of the momentum and summation over the projections
\cite{varshalovich}
we get
\begin{eqnarray}
I_1
&=&\nonumber
\frac{1}{4\pi}
\sum\limits_{Kk}
\Pi(j_1,j_2,j_3,j_4,l_1,l_2,l_3,l_4)
\\
&&\nonumber
\times
(-1)^{l_1+j_2+1/2}
(-1)^{l_3+j_4+1/2}
\\
&&\nonumber
\times
\threej{l_1}{l_2}{K}{0}{0}{0}
\threej{l_3}{l_4}{K}{0}{0}{0}
\sixj{K}{j_2}{j_1}{1/2}{l_1}{l_2}
\sixj{K}{j_4}{j_3}{1/2}{l_3}{l_4}
\\
&&\label{eq1005050320}
\times
C^{j_1j_2}_{Kk}(m_1m_2)
C^{j_3j_4}_{Kk}(m_3m_4)
\,.
\end{eqnarray}
Here, we introduced the $3j$-symbols, $6j$-symbols
\cite{varshalovich}
and
\begin{eqnarray}
\Pi(j_1,j_2,\ldots,j_n)
&=&
\sqrt{2j_1+1}\sqrt{2j_2+1} \ldots \sqrt{2j_n+1}
\,.
\end{eqnarray}

\par
Accordingly,
\Eq{eqn110919n7}
can be written as
\begin{eqnarray}
I
&=&\nonumber
\frac{1}{2}
\left(\frac{(2\pi)^3}{p\oee}\right)^2
\sum\limits_{j_1j_2j_3j_4 l_1l_2l_3l_4 m_1m_2m_3m_4}
\frac{1}{4\pi}
\sum\limits_{Kk}
\Pi(j_1,j_2,j_3,j_4,l_1,l_2,l_3,l_4)
\\
&&\nonumber
\times
\threej{l_1}{l_2}{K}{0}{0}{0}
\threej{l_3}{l_4}{K}{0}{0}{0}
\sixj{K}{j_2}{j_1}{1/2}{l_1}{l_2}
\sixj{K}{j_4}{j_3}{1/2}{l_3}{l_4}
\\
&&\nonumber
\times
(-1)^{l_1+j_2+1/2}
(-1)^{l_3+j_4+1/2}
C^{j_1j_2}_{Kk}(m_1m_2)
C^{j_3j_4}_{Kk}(m_3m_4)
\\
&&\label{eq1005061709}
\times
e^{i\varphi_{\oee j_1l_1} + i\varphi_{\oee j_2l_2} - i\varphi_{\oee j_3l_3} - i\varphi_{\oee j_4l_4}}
\psi_{\oee j_1l_1m_1}(\zhr_1)
\psi_{\oee j_2l_2m_2}(\zhr_2)
\psi^{*}_{\oee j_3l_3m_3}(\zhr_3)
\psi^{*}_{\oee j_4l_4m_4}(\zhr_4)
\,.
\end{eqnarray}
This equation can be written in the form
\begin{eqnarray}
I
&=&\nonumber
\left(\frac{(2\pi)^3}{p\oee}\right)^2
\sum\limits_{Kk}
\sum\limits_{j_1\le j_2, l_1\le l_2, j_3 \le j_4, l_3 \le l_4}
A_{Kk\oee j_ll_1 \oee j_2l_2 \oee j_3l_3 \oee j_4l_4}
\\
&&\nonumber
\times
N_{12}
\sum\limits_{m_1m_2}
C^{j_1j_2}_{Kk}(m_1m_2)
\det\{\psi_{\oee j_1l_1m_1}(\zhr_1)\psi_{\oee j_2l_2m_2}(\zhr_2)\}
\\
&&
\times
N_{34}
\sum\limits_{m_3m_4}
C^{j_3j_4}_{Kk}(m_3m_4)
\det\{\psi^{*}_{\oee j_3l_3m_3}(\zhr_3)\psi^{*}_{\oee j_4l_4m_4}(\zhr_4)\}
\,,
\end{eqnarray}
where $N=1/2$ for equivalent electrons and $N=1/\sqrt{2}$ for non-equivalent electrons; $n=(jlm)$.
Here, we introduced coefficients 
\begin{eqnarray}
A_{Kk\oee j_ll_1 \oee j_2l_2 \oee j_3l_3 \oee j_4l_4}
&=&\nonumber
\frac{1}{4\pi}
\Pi(j_1,j_2,j_3,j_4,l_1,l_2,l_3,l_4)
\\
&&\nonumber
\times
\threej{l_1}{l_2}{K}{0}{0}{0}
\threej{l_3}{l_4}{K}{0}{0}{0}
\sixj{K}{j_2}{j_1}{1/2}{l_1}{l_2}
\sixj{K}{j_4}{j_3}{1/2}{l_3}{l_4}
\\
&&
\times
(-1)^{l_2+j_2+1/2}
(-1)^{l_4+j_4+1/2}
e^{i\varphi_{\oee j_1l_1} + i\varphi_{\oee j_2l_2} - i\varphi_{\oee j_3l_3} - i\varphi_{\oee j_4l_4}}
\,.
\end{eqnarray}

%
%
\begin{table}
\caption{
Radii (in pm) and areas of the cross-sections of the reaction volumes for one incident electron (in kilobarn) for the C target atom
}
\begin{tabular}{c|c|c|c}
\hline
$R^\txt{A}_\txt{T}$
&$S^\txt{A}$
&$R^\txt{K}_\txt{T}$
&$S^\txt{K}$
\\
\hline
$133$&$919$ &$14$&$31$\\
\hline
\end{tabular}
\\
\label{tabradiusC}
\end{table}

%
%
\begin{table}
\caption{
Radii (in pm) and areas of the cross-sections of the reaction volumes for one incident electron (in kilobarn) for the Ar target atom
}
\begin{tabular}{c|c|c|c}
\hline
$R^\txt{A}_\txt{T}$
&$S^\txt{A}$
&$R^\txt{K}_\txt{T}$
&$S^\txt{K}$
\\
\hline
$188$&$617$ &$3.0$&$1.4$\\
\hline
\end{tabular}
\\
\label{tabradiusAr}
\end{table}

%
%
\begin{table}
\caption{
Cross-section (in barn) for RDEC process O${}^{8+}+$C, $\sigma_{(1s1s)}^\txt{RDEC}$ contribution.
}
\begin{tabular}{c|c|c|c}
\hline
Experiment&\multicolumn{3}{|c}{Theory}\\
\hline
$\sigma_{(1s1s)}^\txt{RDEC}$ \cite{simon10phd}
& $\sigma_{(1s1s)}^\txt{RDEC}$ \cite{mikhailov04p350}
&$\sigma_{(1s1s)}^\txt{RDEC,A}$  [this work]
&$\sigma_{(1s1s)}^\txt{RDEC,K}$  [this work]
\\
\hline
 $3.2(1.9)$ & $0.15$ &	$0.55$ & $0.019$ \\
\hline
\end{tabular}
\label{tabO1}
\end{table}

%
%
\begin{table}
\caption{
Cross-section (in barn) for RDEC process O${}^{8+}+$C, $\sigma_{(1s1s)}^\txt{RDEC}$ contribution.
}
\begin{tabular}{c|c|c|c}
\hline
Experiment&\multicolumn{3}{|c}{Theory}\\
\hline
$\sigma_{(1s2s)}^\txt{RDEC}$ \cite{simon10phd}
& $\sigma_{(1s2s)}^\txt{RDEC}$  \cite{nefiodov05}
&$\sigma_{(1s2s)}^\txt{RDEC,A}$ [this work]
&$\sigma_{(1s2s)}^\txt{RDEC,K}$ [this work]
\\
\hline
 $2.2(1.3)$ & $0.105$ &	$0.05$ & $0.002$\\
\hline
\end{tabular}
\\
\label{tabO2}
\end{table}

%
%
\begin{table}
\caption{
Total cross-section (in barn) for RDEC process O${}^{8+}+$C, $\sigma^\txt{RDEC}$ contribution.
}
\begin{tabular}{c|c|c|c}
\hline
Experiment&\multicolumn{3}{|c}{Theory}\\
\hline
$\sigma^\txt{RDEC}$ \cite{simon10phd}
& $\sigma^\txt{RDEC}$ \cite{mikhailov04p350,nefiodov05}
&$\sigma^\txt{RDEC,A}$ [this work]
&$\sigma^\txt{RDEC,K}$ [this work]
\\
\hline
 $5.5(3.2)$ & $0.26$ &	$0.61$ & $0.021$ \\
\hline
\end{tabular}
\label{tabO3}
\end{table}

%
%
\begin{table}
\caption{
Cross-section (in millibarn) for RDEC process Ar${}^{18+}+$C, $\sigma_{(1s1s)}^\txt{RDEC}$ contribution.
}
\begin{tabular}{c|c|c|c}
\hline
Experiment&\multicolumn{3}{|c}{Theory}\\
\hline
$\sigma_{(1s1s)}^\txt{RDEC}$ \cite{warczak95}
& $\sigma_{(1s1s)}^\txt{RDEC}$ \cite{mikhailov04p350} 
&$\sigma_{(1s1s)}^\txt{RDEC,A}$ [this work]
&$\sigma_{(1s1s)}^\txt{RDEC,K}$ [this work]
\\
\hline
 $\le 5.2$ & $3.2$ & $120$ & $4.3$\\
\hline
\end{tabular}
\label{tabAr1}
\end{table}

%
%
\begin{table}
\caption{
Cross-section (in millibarn) for RDEC process U${}^{92+}+$Ar, $\sigma_{(1s1s)}^\txt{RDEC}$ contribution.
}
\begin{tabular}{c|c|c|c}
\hline
Experiment&\multicolumn{3}{|c}{Theory}\\
\hline
$\sigma_{(1s1s)}^\txt{RDEC}$ \cite{bednarz03}
& $\sigma_{(1s1s)}^\txt{RDEC}$ \cite{mikhailov04p350} 
&$\sigma_{(1s1s)}^\txt{RDEC,A}$ [this work]
&$\sigma_{(1s1s)}^\txt{RDEC,K}$ [this work]
\\
\hline
 $< 10$ & $2.5\times10^{-2}$&	$1.73$&	$0.31\times10^{-2}$\\
\hline
\end{tabular}
\\
\label{tabU1}
\end{table}

%
%

%
%

%
%
%

%
\end{document}